\documentclass[11pt]{article}
\usepackage{moriond,epsfig}

\def\Journal#1#2#3#4{{#1} {\bf #2}, #3 (#4)}

\def\PRL{\em Phys. Rev. Lett.}
\def\PRD{{\em Phys. Rev.} D}

\def\be{\begin{equation}}
\def\ee{\end{equation}}
\def\bea{\begin{eqnarray}}
\def\eea{\end{eqnarray}}

\def\be{\beta}

\def\th{\theta}

\def\si{\sigma}

\def\De{\Delta}

\def\Ph{\Phi}

\def\fr#1#2{{{#1} \over {#2}}}
\def\frac#1#2{\textstyle{{{#1} \over {#2}}}}

\begin{document}
\vspace*{4cm}
\title{NEUTRINO CROSS SECTION MEASUREMENTS FOR LONG-BASELINE 
ACCELERATOR-BASED NEUTRINO OSCILLATION EXPERIMENTS}

\author{TEPPEI KATORI}

\address{Indiana University, Bloomington, Indiana, USA}

\maketitle\abstracts{
Neutrino oscillations are clear evidence for physics beyond the standard model. 
The goal of next-generation neutrino oscillation experiments 
is to find a non-zero $\theta_{13}$, 
the last mixing matrix element for which we only know an upper limit.
For this, next-generation long-baseline neutrino oscillation experiments 
require an order of magnitude better sensitivities. 
In particular, accelerator-based experiments such as T2K and NOvA experiments need 
(1) good neutrino energy reconstruction for the precise measurement of $\Delta m^2_{32}$ 
and $sin^22\theta_{23}$, and (2) good background prediction to measure $\nu_e$ 
appearance signals. Current and near future high statistics 
neutrino experiments, 
such as K2K, MiniBooNE, SciBooNE, MINOS, and MINERvA help both (1) and (2) 
by precise signal and background channel measurements.  
}

\section{next-generation long baseline accelerator-based neutrino oscillation experiments}

The goal of next-generation long baseline accelerator-based neutrino 
oscillation experiments is 
to measure a non-zero $\theta_{13}$, the last mixing matrix element. 
The value of $\th_{13}$ is the important parameter to access beyond the standard model physics. 
Especially if it were non-zero, then we hope to measure leptonic CP violation 
which can help to understand leptogenesis, 
one of the candidate explanations of baryon asymmetry of the universe~\cite{theory}.  

Currently two experiments are planned, 
the Tokai-to-Kamioka (T2K) experiment~\cite{T2K} ($\sim 800$ MeV, $\sim 300$ km) and 
the NuMI Off-axis $\nu_e$ Appearance (NOvA) experiment~\cite{NOvA} 
($\sim 2$ GeV, $\sim 800$ km). 
Both experiments use a $\nu_\mu$ beam and search for $\nu_e$ appearance events 
to measure $\theta_{13}$ through the equation, 
\begin{eqnarray}
P(\nu_\mu\to\nu_e) = sin^2\theta_{23}sin^22\theta_{13}
sin^2\left(1.27\fr{\De m_{32}^2(eV^2)L(km)}{E(GeV)}\right).
\end{eqnarray}

\noindent
Since a small $P(\nu_\mu\to\nu_e)$ is proportional to 
$sin^2\theta_{23}$ and $sin^2\left(1.27\frac{\De m_{32}^2L}{E}\right)$, 
we also need accurate knowledge of these two quantities, 
and can achieve by the measurements of $\nu_\mu$ disappearance events,
\begin{eqnarray}
P(\nu_\mu\to\nu_\mu)=1-sin^22\theta_{23}
sin^2\left(1.27\fr{\De m_{32}^2(eV^2)L(km)}{E(GeV)}\right).
\end{eqnarray}

\noindent
The oscillation parameters are extracted from the shape of $P(\nu_\mu\to\nu_\mu)$, 
a function of reconstructed neutrino energy. 
Therefore a good extraction of $sin^2\theta_{23}$ and $\De m_{32}^2$ 
rely on good reconstruction of neutrino energy, which is based on 
better understanding of the signal ($\nu_\mu$CCQE) and background interactions, 
mainly CC1$\pi^{\circ}$ interaction (Sec.~\ref{sec:recon}). 

\vspace{0.5cm}
The signal of $\nu_e$ appearance is an electron,
\begin{eqnarray}
\nu_e+n \to p+e^-.
\label{eq:nueCCQE}
\end{eqnarray}

\noindent
There are many kind of possible backgrounds for this signal, 
for example, sometimes $\nu_\mu$ induced NC$\pi^{\circ}$ production can mimic a $\nu_e$ event 
if one of the decay photons from $\pi^{\circ}$ decay is undetected. 
Therefore, it is critical to understand this background channel (Sec.~\ref{sec:bkgd}). 

\begin{figure}
\vskip 0.0cm
\hskip 3.0cm
\includegraphics[height=3.0in]{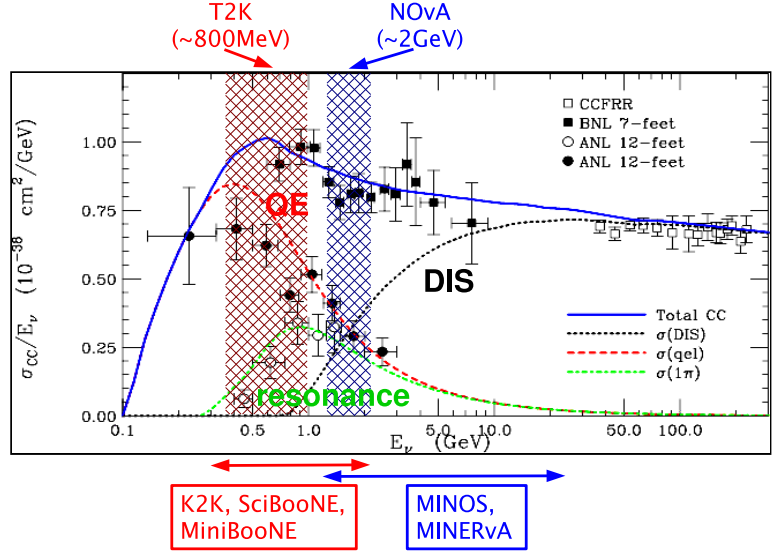}
\caption{The world data for $\nu_\mu$ charged current cross section 
divided by neutrino energy. 
The dominant interaction for T2K and NOvA are quasi-elastic (QE)  
and deep inelastic scattering (DIS) respectively. 
The existing data are rather sparse and old, but we have more new input from 
current and future experiments!
}
\label{fig:cc_all}
\end{figure}

\vspace{0.5cm}
It is important to perform these cross section measurements prior to 
oscillation experiments. Although all long baseline accelerator-based 
neutrino oscillation experiments have near detectors, 
they exist to constrain neutrino flux uncertainties, 
and this constraint relies on accurate knowledge of cross section measurements. 
Fig.~\ref{fig:cc_all} shows the world's data for charged current cross sections. 
As you can see, existing data are rather sparse and old. 
Since two experiments, T2K and NOvA, 
span different energy ranges, we need cross section measurements in both regions 
because the dominant interaction types will be different, 
and thus their energy reconstructions and backgrounds are different. 
Fortunately, we have a lot of new input from current and future neutrino cross section measurements: 
K2K near detector~\cite{K2K_ccqe} ($\sim 1.2$ GeV, completed), 
MiniBooNE~\cite{MB_ccqe} ($\sim 800$ MeV, ongoing), 
SciBooNE~\cite{SciBooNE} ($\sim 800$ MeV, ongoing), 
MINOS near detector~\cite{MINOS} ($\sim 2-20$ GeV, ongoing), and 
MINERvA~\cite{MINERvA} ($\sim 2-20$ GeV, approved). 
We would like to discuss the two main themes of cross section related issues 
impacting oscillation searches, 
(1) neutrino energy reconstruction (Sec.~\ref{sec:recon}), and 
(2) background determination (Sec.~\ref{sec:bkgd}).

\section{Neutrino energy reconstruction \label{sec:recon}}

\subsection{Neutrino energy reconstruction for T2K}

At the T2K energy scale ($\sim 800$ MeV), the dominant neutrino reactions are 
$\nu_\mu$ charged current quasi-elastic (CCQE) interactions, 
\begin{eqnarray}
\nu_\mu+n \to p+\mu^-.
\label{eq:numuCCQE}
\end{eqnarray}

\noindent
This channel is used to measure $\nu_\mu$ disappearance, 
and thus the $\nu_\mu$ energy reconstruction is critical. 
Since neutrino oscillation experiments use nuclear targets, 
understanding of this interaction is not trivial. 
Recently K2K~\cite{K2K_ccqe,K2K_ccqe2} and MiniBooNE~\cite{MB_ccqe} have reported 
new measurements of the axial mass, $M_A$, which are higher than the historical value 
(Table~\ref{tab:ma}). 

\begin{table}
\begin{center}
\begin{tabular}{|l|c|c|}
\hline
              & $M_A$(GeV)        & target      \\
\hline
K2K (SciFi)~\cite{K2K_ccqe}  & 1.20  $\pm$ 0.12  & oxygen        \\
K2K (SciBar)~\cite{K2K_ccqe2}& 1.14  $\pm$ 0.11  & carbon        \\
MiniBooNE~\cite{MB_ccqe}     & 1.23  $\pm$ 0.20  & carbon        \\
world average~\cite{WA_ccqe} & 1.026 $\pm$ 0.021 & deuteron, etc \\
\hline
\end{tabular}
\end{center}
\caption{The comparison of measured axial mass $M_A$.}
\label{tab:ma}
\end{table}

\begin{figure}
\vskip 0.0cm
\hskip 1.0cm
\includegraphics[height=2.0in]{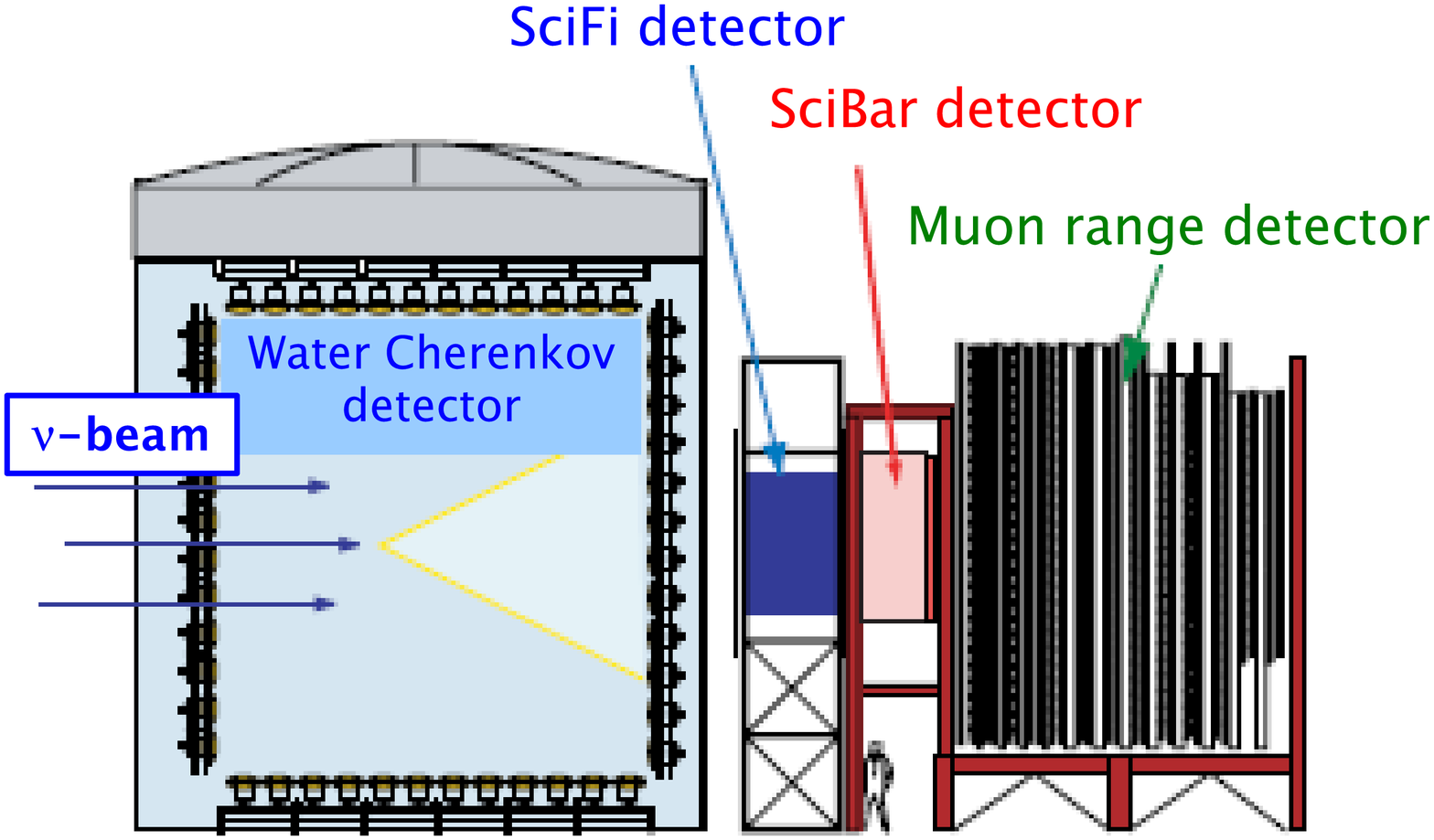}
\hskip 0.5cm
\includegraphics[height=2.0in]{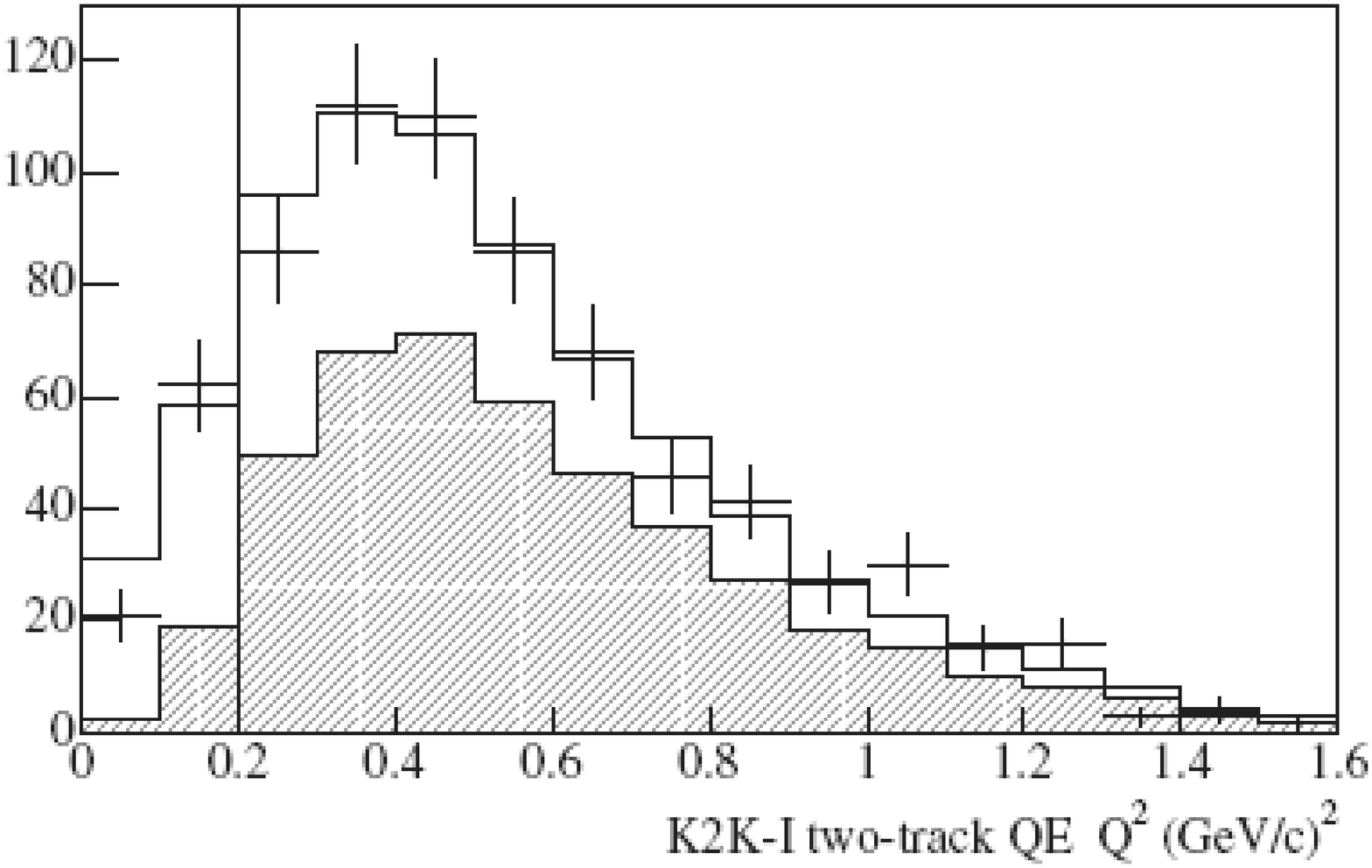}
\caption{
(Left) K2K near detector complex. From the left to right, 
1 kiloton water \v{C}erenkov detector ``1KT'', 
scintillation-fiber/water target tracker ``SciFi'', 
fully active plastic organic scintillation-bar tracker ``SciBar'', and 
muon range detector ``MRD''. 
(Right) reconstructed $Q^2$ plot for 2-track QE sample from K2K SciFi, 
data (crosses) and simulation with best-fit $M_A$ (solid) agree well. 
The shaded region indicates the fraction of signal ($\nu_\mu CCQE$) events. 
}
\label{fig:K2K_ccqe}
\end{figure}

\begin{figure}
\vskip 0.0cm
\hskip 0.0cm
\includegraphics[height=1.5in]{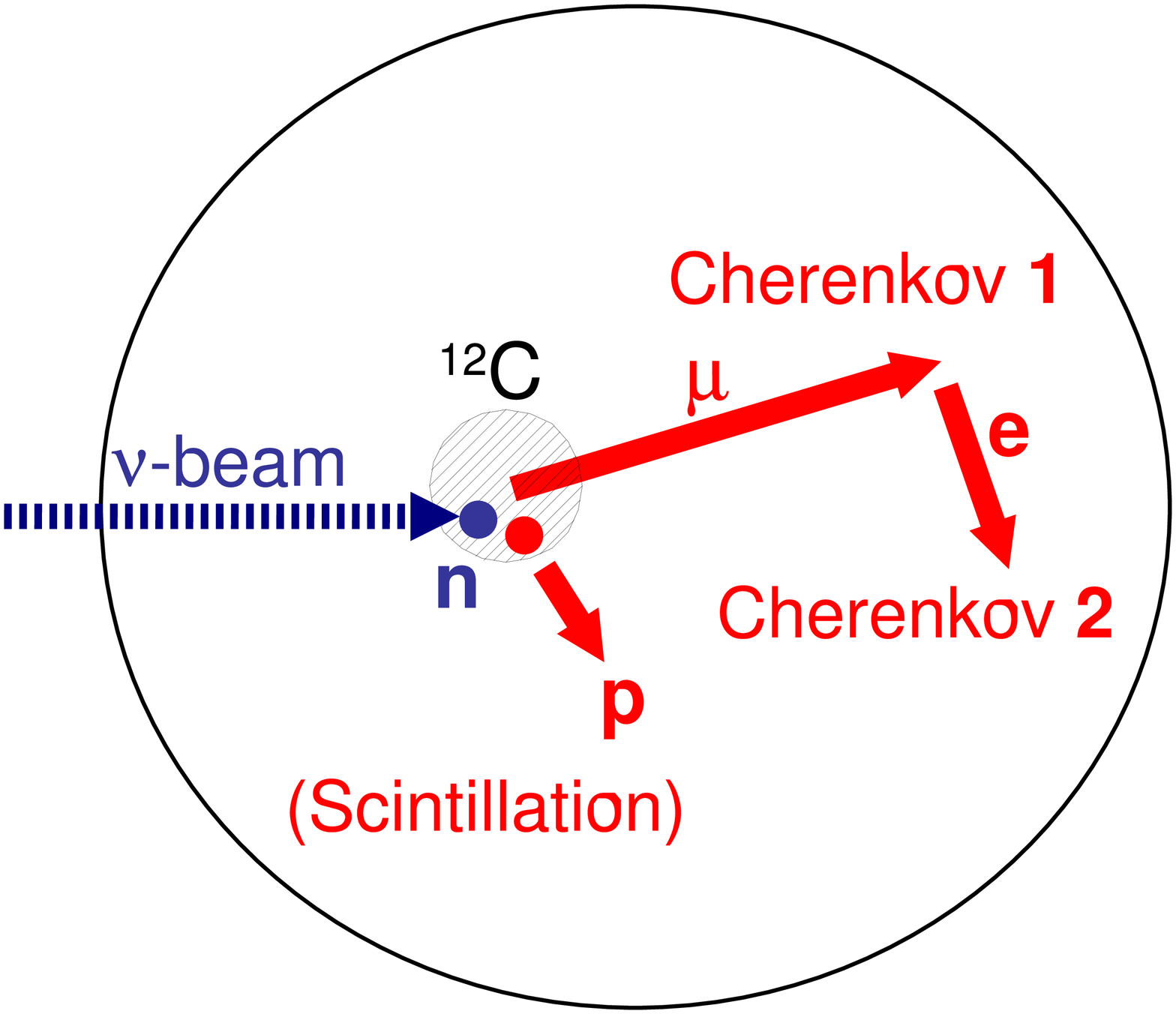}
\includegraphics[height=1.5in]{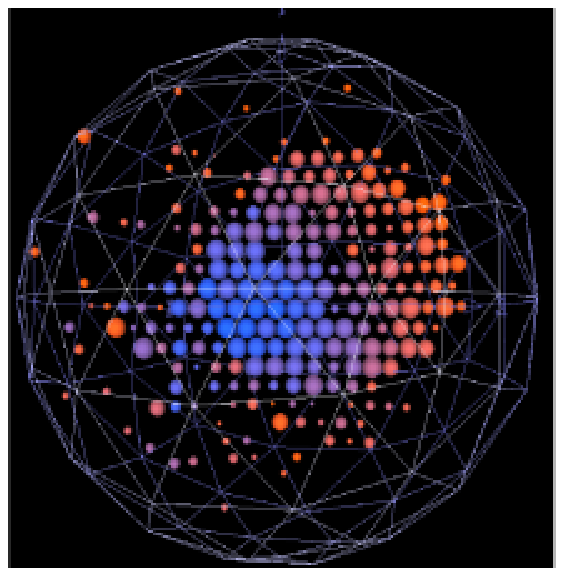}
\includegraphics[height=1.5in]{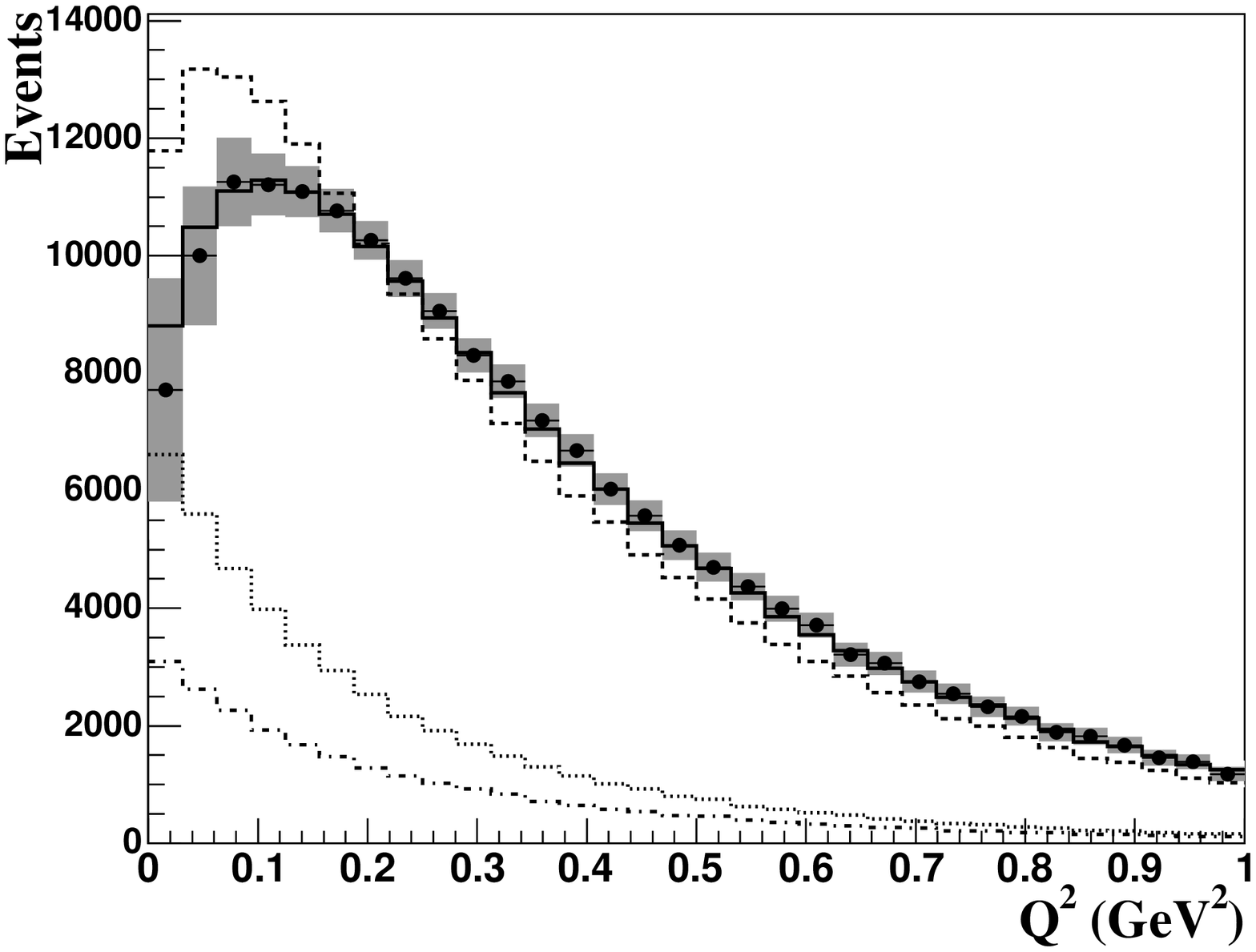}
\caption{
(Left) Schematic figure of a $\nu_\mu$ CCQE interaction in MiniBooNE. 
The MiniBooNE detector is a \v{C}erenkov detector filled with mineral oil surrounded by PMTs.  
The \v{C}erenkov light from the muon (Cherenkov 1) and subsequent \v{C}erenkov light from 
the decayed electron are used to tag the CCQE event. 
(Middle) Event display of a muon candidate event in MiniBooNE. Each sphere represents a hit on a PMT, 
and size and color show charge and time information respectively.
Muons create shape-edged \v{C}erenkov ring. 
The ring center will appear filled-in if the muon is stopping in the tank.  
(Right) Reconstructed $Q^2$ plot of MiniBooNE, 
data (dots), simulation before the fit (dashed), 
and after the fit with $M_A$ and Pauli-blocking (solid). 
The dotted and dash-dotted lines indicate total 
background and irreducible background fraction respectively.}
\label{fig:MB_ccqe}
\end{figure}

In this energy range, the axial vector form factor is the dominant contribution 
to the cross section and controls the $Q^2$ dependence. 
Inconsistency of their results from the world average, 
and the consistency between K2K and MiniBooNE is best 
understood in terms of nuclear effects, 
because most of the past experiments used deuterium targets whereas 
K2K and MiniBooNE used oxygen and carbon. 
Instead of using the world average, 
both experiments employ their measured $M_A$ values 
to better simulate CCQE events in their oscillation analyses. 
After the $M_A$ adjustment, both experiments see good agreement 
between data and simulation (Fig~\ref{fig:K2K_ccqe} and \ref{fig:MB_ccqe}). 

\begin{figure}
\vskip 0.0cm
\hskip 0.5cm
\includegraphics[height=1.6in]{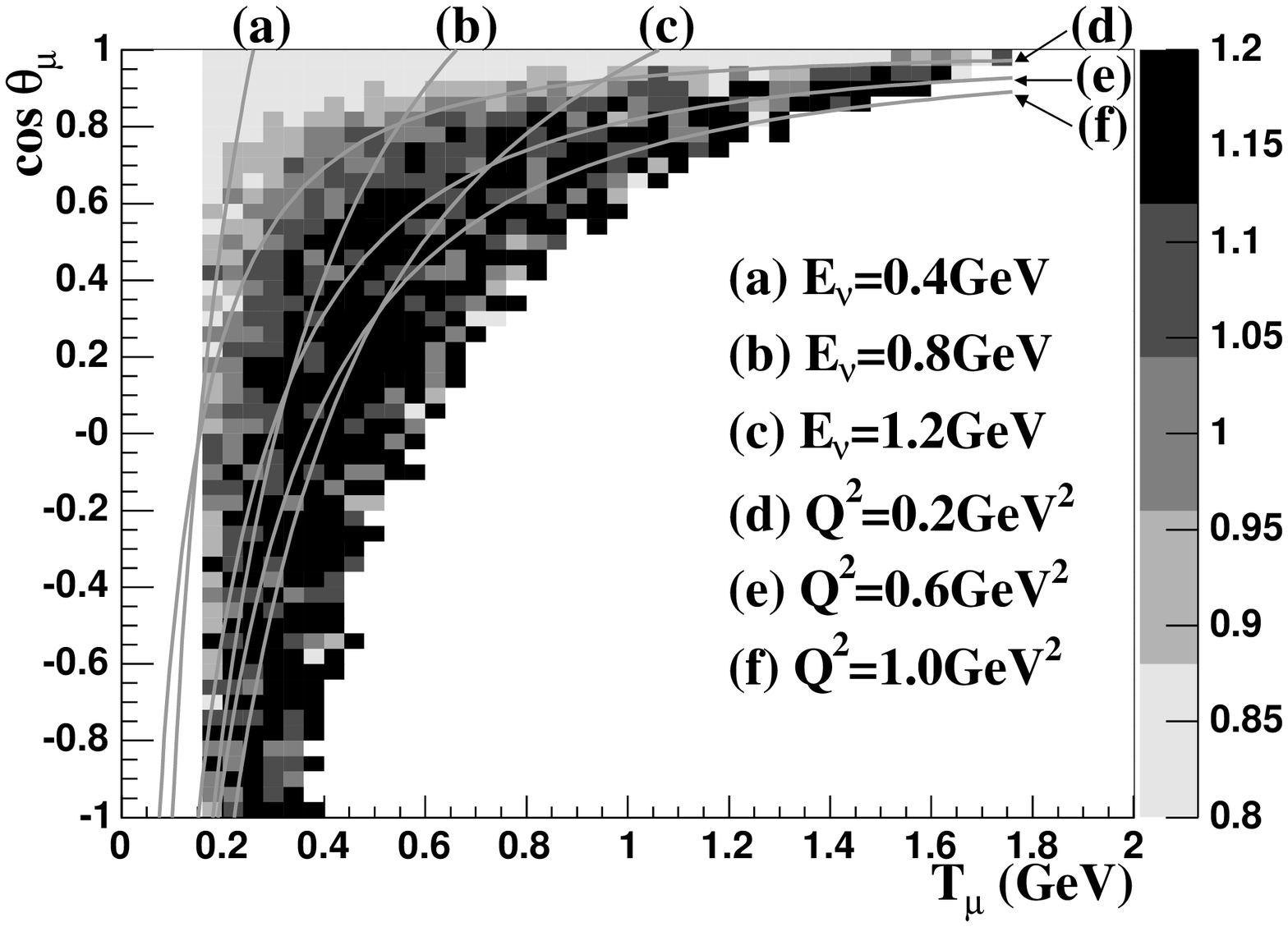}
\includegraphics[height=1.6in]{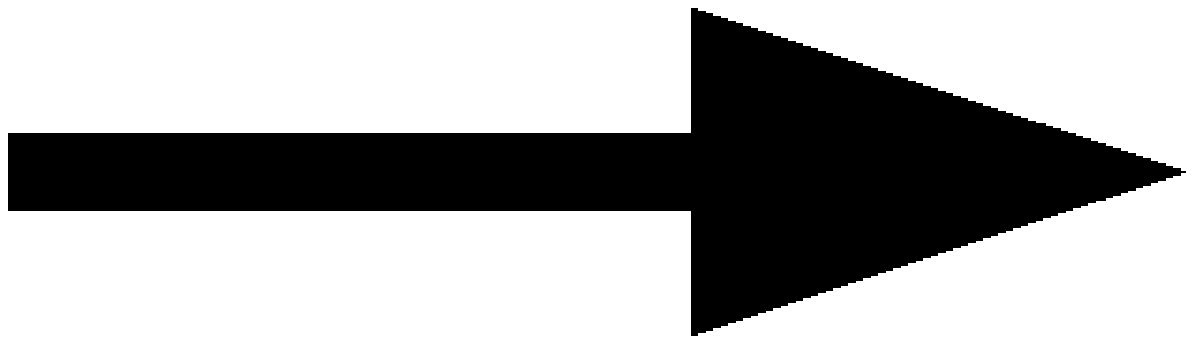}
\includegraphics[height=1.6in]{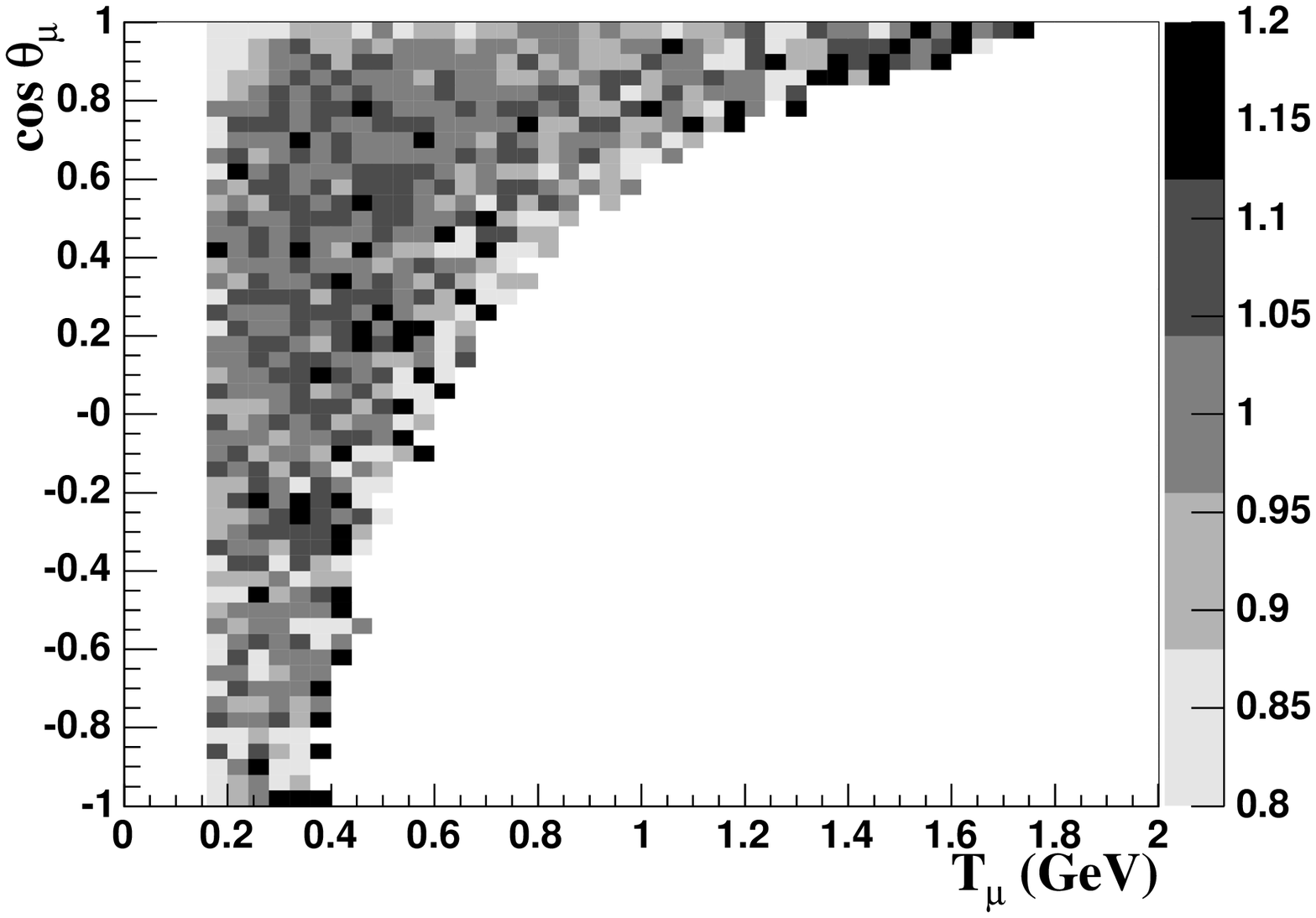}
\caption{
Ratio of MiniBooNE $\nu_\mu$CCQE data-simulation 
in the 2-dimensional plane of muon kinetic energy vs. muon angle. 
If the prediction well-describes the data, 
then this plot should exhibit a uniform distribution centered on unity. 
(Left) before CCQE cross section model tuning, 
the light gray region shows an excess of predicted events, 
and the black region shows a deficit of predicted events. 
The auxiliary lines from (a) to (f) indicate lines of equal $E_{\nu}$ or $Q^2$. 
The data-simulation discrepancy follows line of constant $Q^2$, 
suggesting an incorrect cross section model in the simulation. 
(Right) after cross section model tuning, specifically adjustment of $M_A$ and Pauli-blocking.
}
\label{fig:ccqe_2dim}
\end{figure}

\hspace{0.5cm} 
We can only measure the interaction rate, 
which is the convolution of flux and cross section ($R=\int\Ph\times\si$). 
So, without knowing flux prediction is perfect, 
one cannot tune the cross section model from measured interaction rate.  
MiniBooNE carefully examined this, and showed that their observed data simulation mismatching 
is not the effect of mismodeling of neutrino flux, but is really a cross section model problem. 
Fig~\ref{fig:ccqe_2dim} shows the ratio of data-simulation 
in the 2-dimensional plane made in muon kinetic energy and angle; 
left plot is before any cross section model tuning, right plot is after. 
The key point is that left plot clearly shows that data-simulation disagreements 
follow equal $Q^2$ lines, not equal $E_{\nu}$ lines. 
\begin{eqnarray}
R = \int \Ph \times \si~\to~R[E_\nu,Q^2] = \int \Ph[E_\nu] \times \si[Q^2]
\end{eqnarray}

\noindent
This is strong evidence that the MiniBooNE data suggests a problem with the cross section model, 
and not the beam model, because cross section is the function of $Q^2$, 
whereas neutrino beam is a function of $E_{\nu}$.

\vspace{0.5cm}
It is not only important to understand the energy reconstruction of signal events 
({\it i.e.}, CCQE interaction), but also for background channels. 
For Super-K, the neutrino energy is reconstructed from the measured 
muon energy $E_{\mu}$ and angle $\theta_{\mu}$, 
assuming a CCQE interaction,
\begin{eqnarray}
E_{\nu}^{QE} \sim
\fr{M_NE_{\mu}-\frac{1}{2}m_{\mu}^2}
{M_N-E_{\mu}+\sqrt{E_{\mu}^2-m_{\mu}^2}cos\th_\mu}.
\label{eq:recon_qe}
\end{eqnarray}

\begin{figure}
\vskip 0.0cm
\hskip 1.0cm
\includegraphics[height=2.0in]{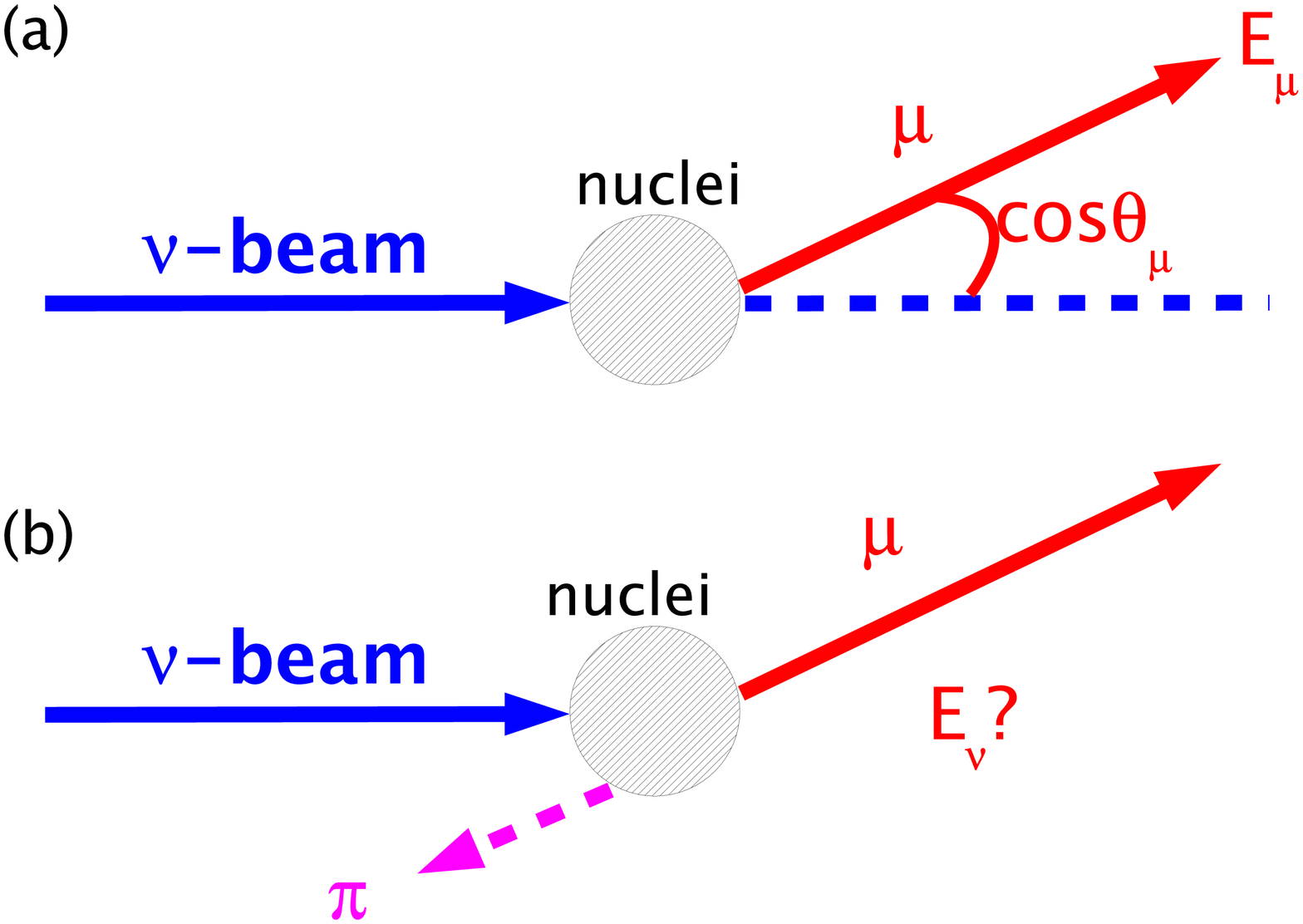}
\includegraphics[height=2.0in]{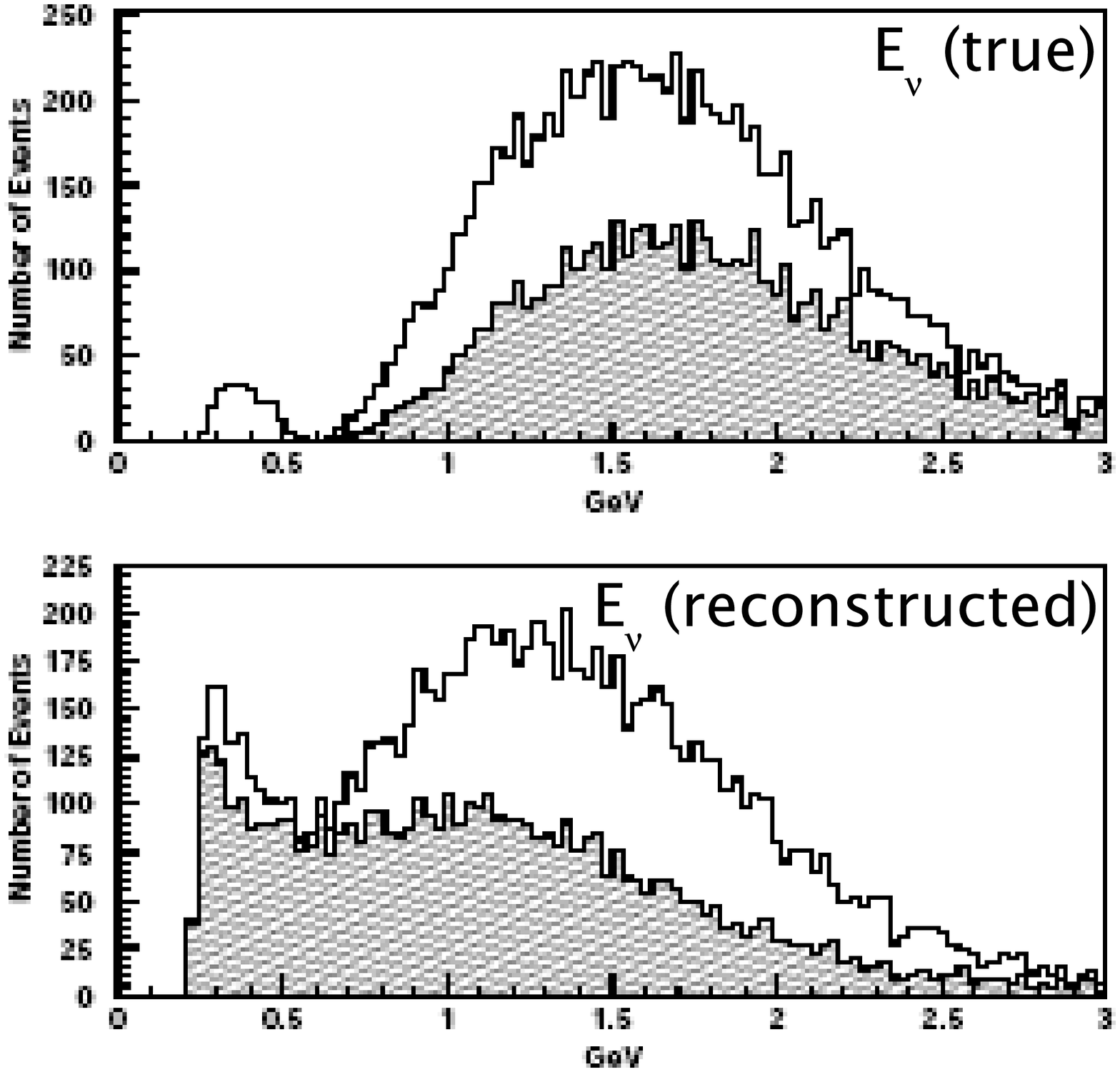}
\caption{
(Left) (a) CCQE interaction and (b) CC1$\pi$ interaction. 
Eq.~\ref{eq:recon_qe} correctly reconstructs neutrino energy only for (a). 
(b) can be distinguished from (a) by additional pion, 
however when pion is lost (by pion absorption for example), 
(b) becomes indistinguishable from intrinsic backgrounds. 
When (a) and (b) have the same muon kinematics, 
the reconstructed neutrino energies are the same, 
however the true neutrino energy for (b) is higher 
due to the creation of the pion in the event (neutrino energy mis-reconstruction).
(Right) true and reconstructed neutrino energy distribution for Super-K predictions 
with neutrino oscillations. 
The shaded region is non-QE (mainly CC1$\pi$) channels. 
As can be seen from the bottom plot, 
CC1$\pi$ background events are misreconstructed at lower neutrino energies 
and hence can fill out the dip created by neutrino oscillations.    
}
\label{fig:K2K_ccpi}
\end{figure}

\noindent
Here, $M_N$ and $m_{\mu}$ are nucleon and muon masses. 
Since this formula assumes a 2-body interaction, 
any interaction involving more than two particles is a source of neutrino 
energy mis-reconstruction (Fig~\ref{fig:K2K_ccpi}, left).  
The most notable channel contributing to this is charged current 1 $\pi$ (CC1$\pi$) production. 
Especially when the detection of the outgoing pion fails for various reasons 
(pion absorption, detector effect, etc), 
CC1$\pi$ events become an irreducible background, 
and thus they need to understand their relative contribution rather than rejecting them by 
cuts~\cite{CWalter} (Fig.~\ref{fig:K2K_ccpi}, right). 

\begin{figure}
\vskip 0.0cm
\hskip 1.0cm
\includegraphics[height=1.8in]{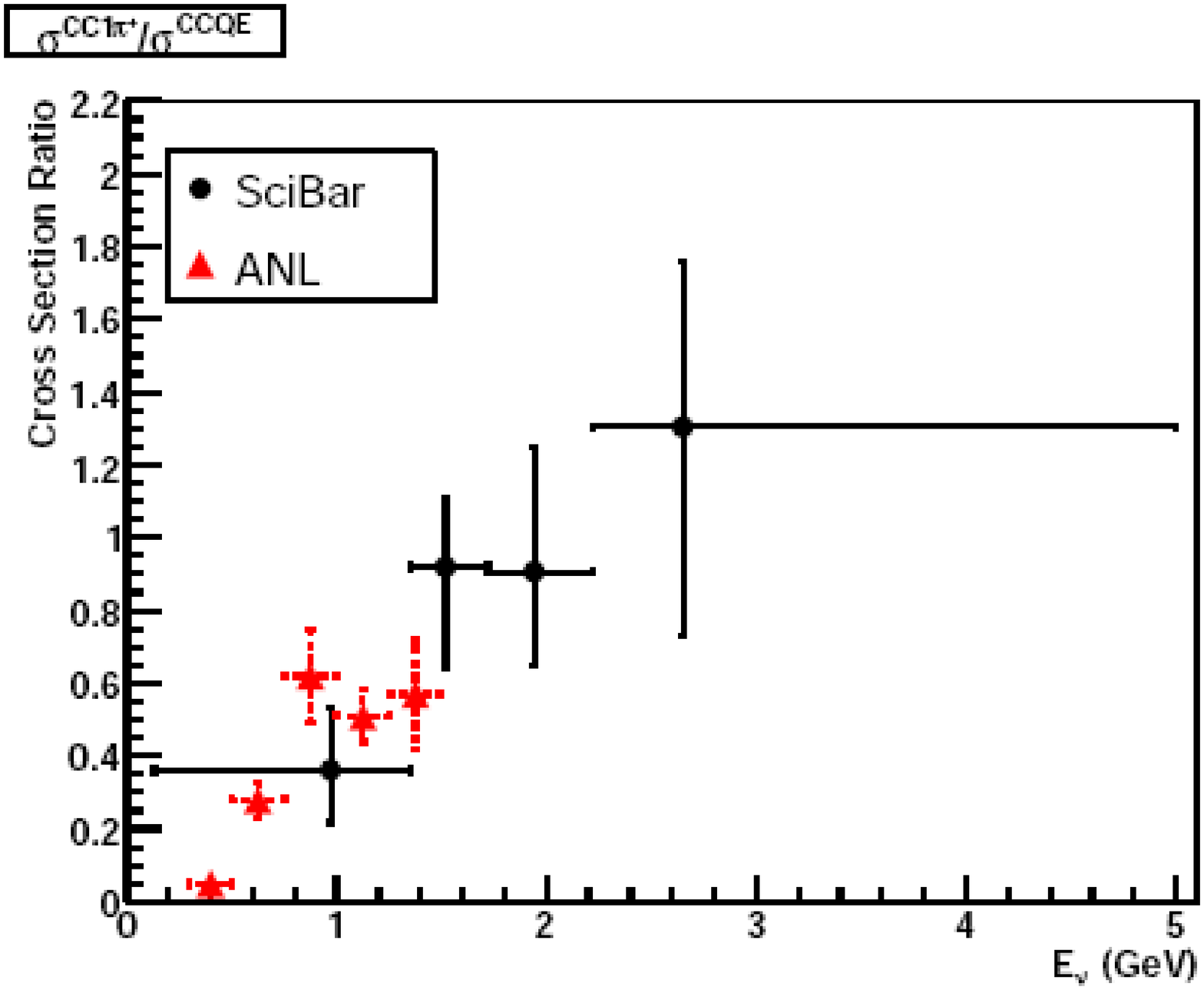}
\hskip 1.0cm
\includegraphics[height=1.8in]{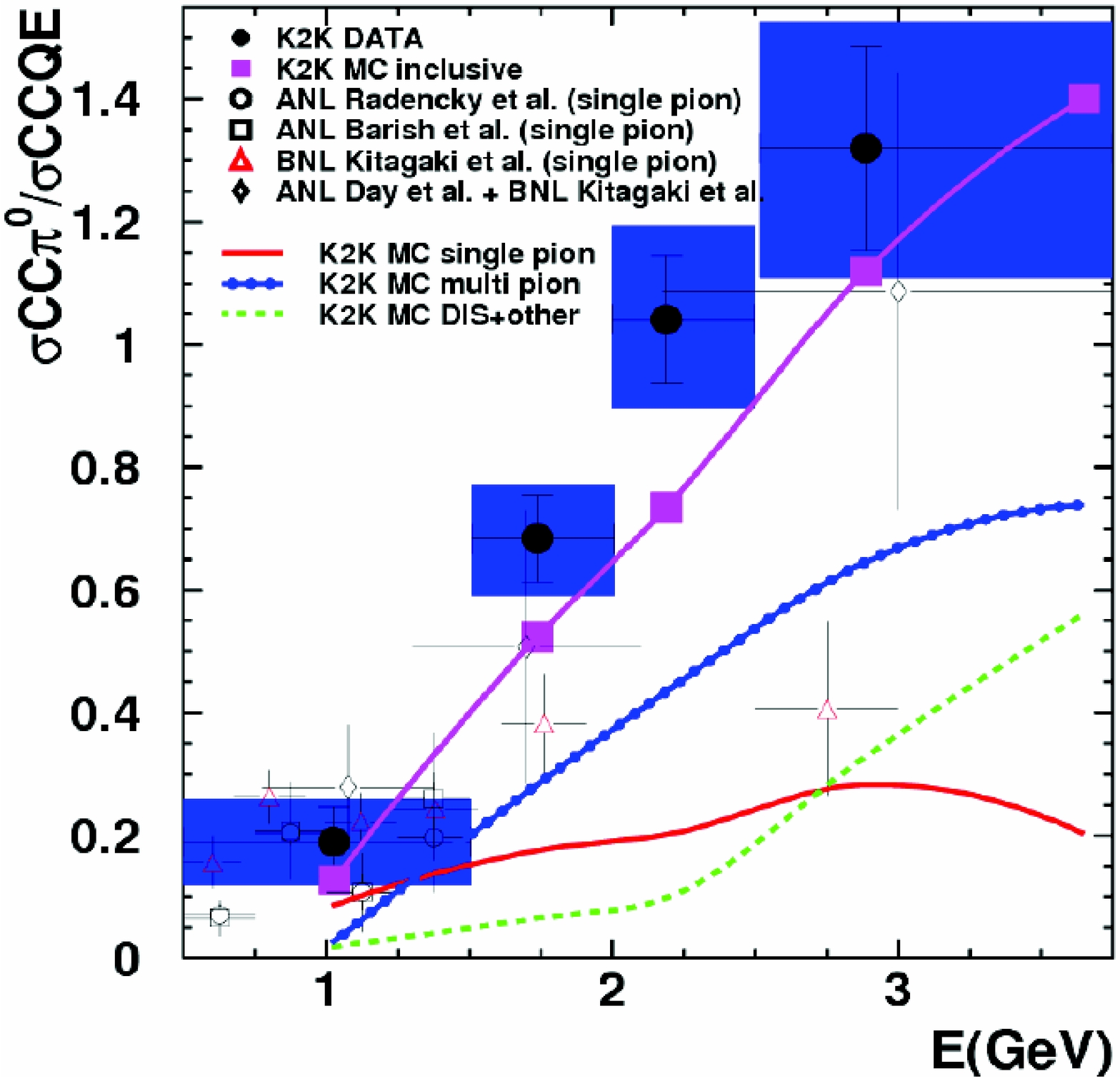}
\caption{
(Left) charged current 1$\pi$ production to CCQE cross section ratio from K2K SciBar analysis. 
Their result is consistent with past ANL bubble chamber experiment. 
(Right) charged current inclusive 1$\pi^{\circ}$ production to CCQE cross section ratio 
from K2K SciBar analysis. 
Although the errors are large, the cross section obtained is 
significantly higher than the cross section model used in the K2K experiment. 
}
\label{fig:K2K_cc1pi}
\end{figure}

Although neutrino absolute cross sections are notoriously difficult to measure 
due to uncertainties in the incoming neutrino flux, 
here they only need to know the kinematic distribution of CC1$\pi$ events 
compared with CCQE events. Such measurements were done in 
K2K (Fig.~\ref{fig:K2K_cc1pi})~\cite{K2K_ccpip,K2K_ccpip2} and MiniBooNE~\cite{MB_ccpip}.

\begin{figure}
\vskip 0.0cm
\hskip 1.0cm
\includegraphics[height=1.5in]{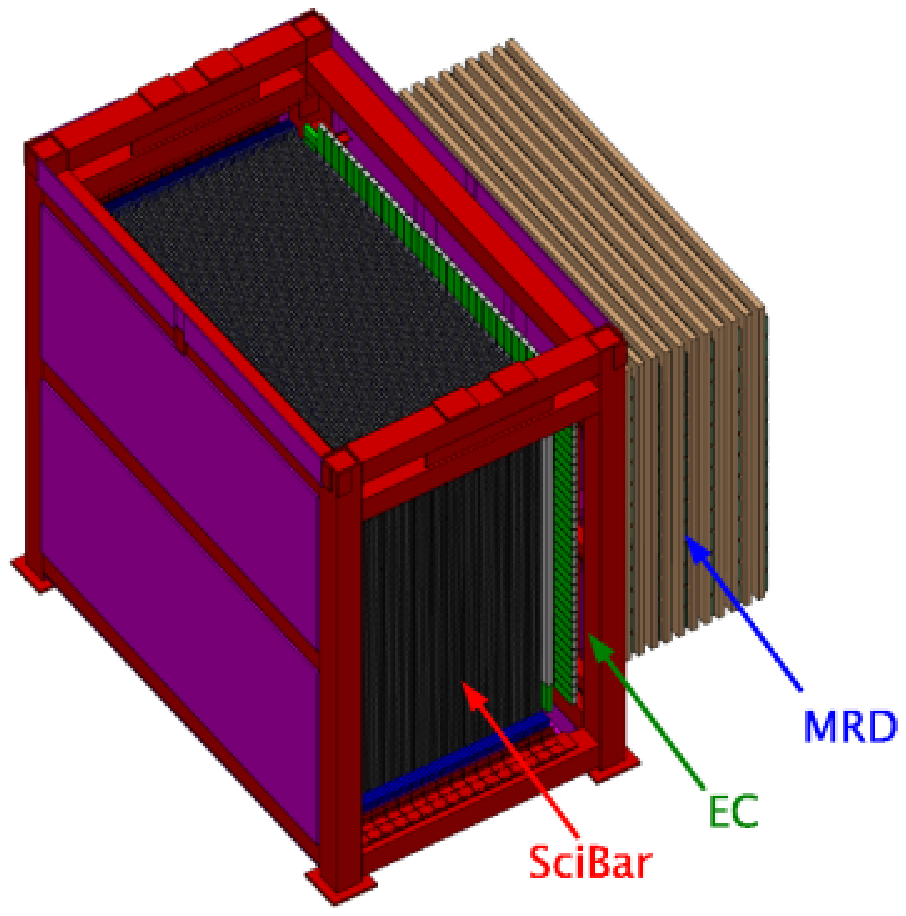}
\includegraphics[height=1.5in]{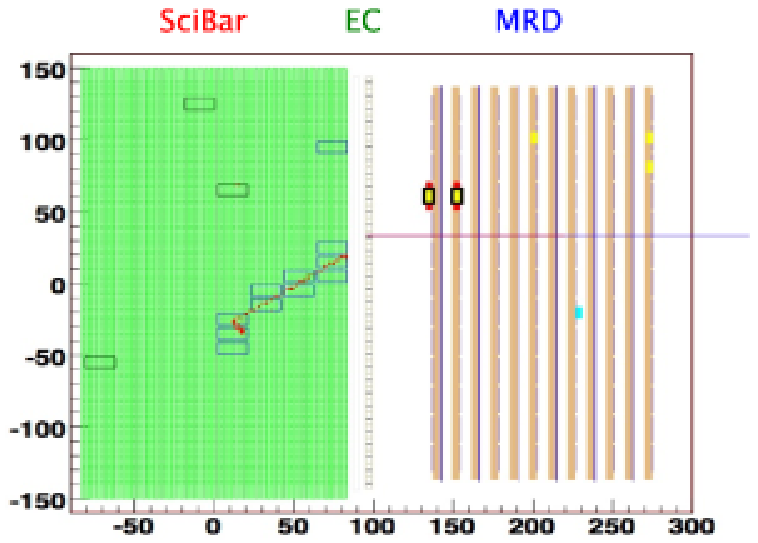}
\includegraphics[height=1.5in]{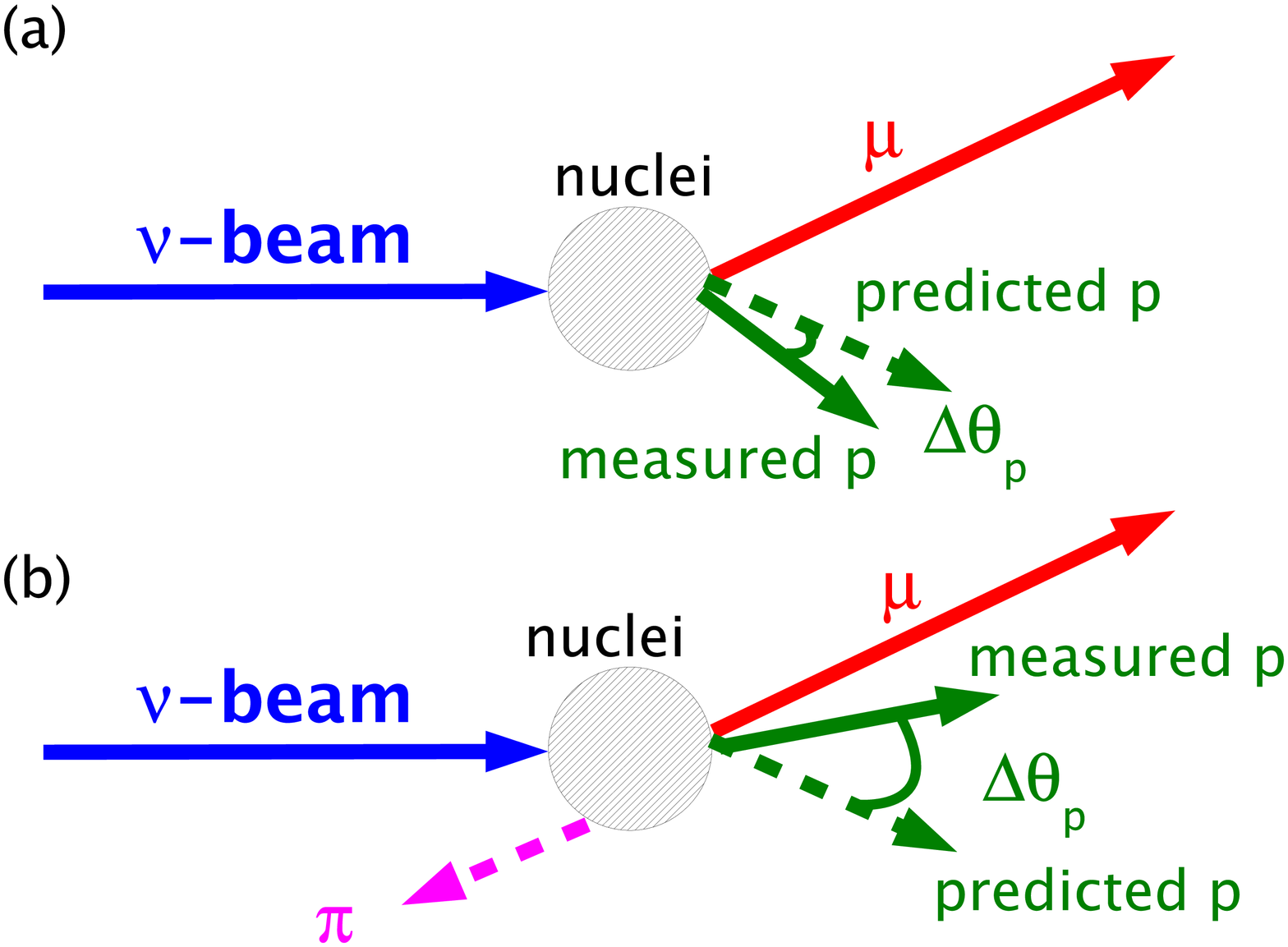}
\caption{
(Left) SciBooNE detector. 
It consists of 3 parts, organic plastic scintillation-bar tracker ``SciBar'', 
11 radiation length lead electromagnetic calorimeter ``EC'', 
and muon range detector ``MRD'' which can range out muons up to $0.9$ GeV.  
(Middle) SciBooNE event display for $\nu_\mu$ CCQE candidate event.
two tracks are seen in ``SciBar'', then the longer track (muon) produce hits in both ``EC'' and ``MRD''.
(Right) Under the assumption of target nucleon at rest, muon energy and angle completely specify 
CCQE kinematics, {\it i.e.}, one can predict the angle of outgoing proton. 
$\De\th_p$ is defined as an opening angle of this predicted proton track and measured proton track. 
(a) is the case of CCQE interaction, and $\De\th_p$ is small. However, 
(b) CC1$\pi$ interaction with invisible pion, 
$\De\th_p$ is large because predicted track is based on the assumption of 2-body interaction 
but actual interaction is 3-body.
}
\label{fig:ccpip}
\end{figure}

The SciBooNE experiment~\cite{SciBooNE} 
at FNAL is particularly designed for this purpose (Fig.~\ref{fig:ccpip}, left and middle). 
The SciBooNE vertex detector ``SciBar'', formerly used at K2K experiment and  
shipped from Japan to Fermilab, 
is a high resolution tracker consisting of X-Y 
plastic organic scintillators with 
wavelength shifting fibers through the middle of each bar.  
Since SciBar can reconstruct both proton and muon tracks in a $\nu_\mu$ CCQE interaction 
(unlike \v{C}erenkov detectors), 
so the opening angle of the measured proton and the expected outgoing proton (assuming CCQE kinematics) 
can be used to separate CCQE and CC1$\pi$ events, 
even in cases where the pion is undetected (right plot of Fig.~\ref{fig:ccpip}). 
The goal of the SciBooNE experiment is to measure non-QE to CCQE cross section ratio to 5\%, 
making the non-QE mis-reconstruction uncertainty for T2K negligible~\cite{SciBooNE}.

\subsection{Neutrino energy reconstruction for NOvA and MINOS}

The situation is quite different for higher energy scales ($\sim 2$ GeV). 
The CCQE assumption is no longer held 
and calorimetric energy reconstruction provides a much more efficient energy determination:
\begin{eqnarray}
E_{\nu} \sim E_{\mu}+E_{showers}
\end{eqnarray}

\noindent
Here, $E_{\mu}$ is the energy of muon, usually measured by a muon spectrometer 
which consists of a dense material to stop muons. 
$E_{showers}$ is the energy of 
both electromagnetic and hadronic showers measured in the calorimeter. 
This energy reconstruction method is successfully tested 
by the Main Injector Neutrino Oscillation Search (MINOS) experiment~\cite{MINOS}. 

\begin{figure}
\vskip 0.0cm
\hskip 1.0cm
\includegraphics[height=1.5in]{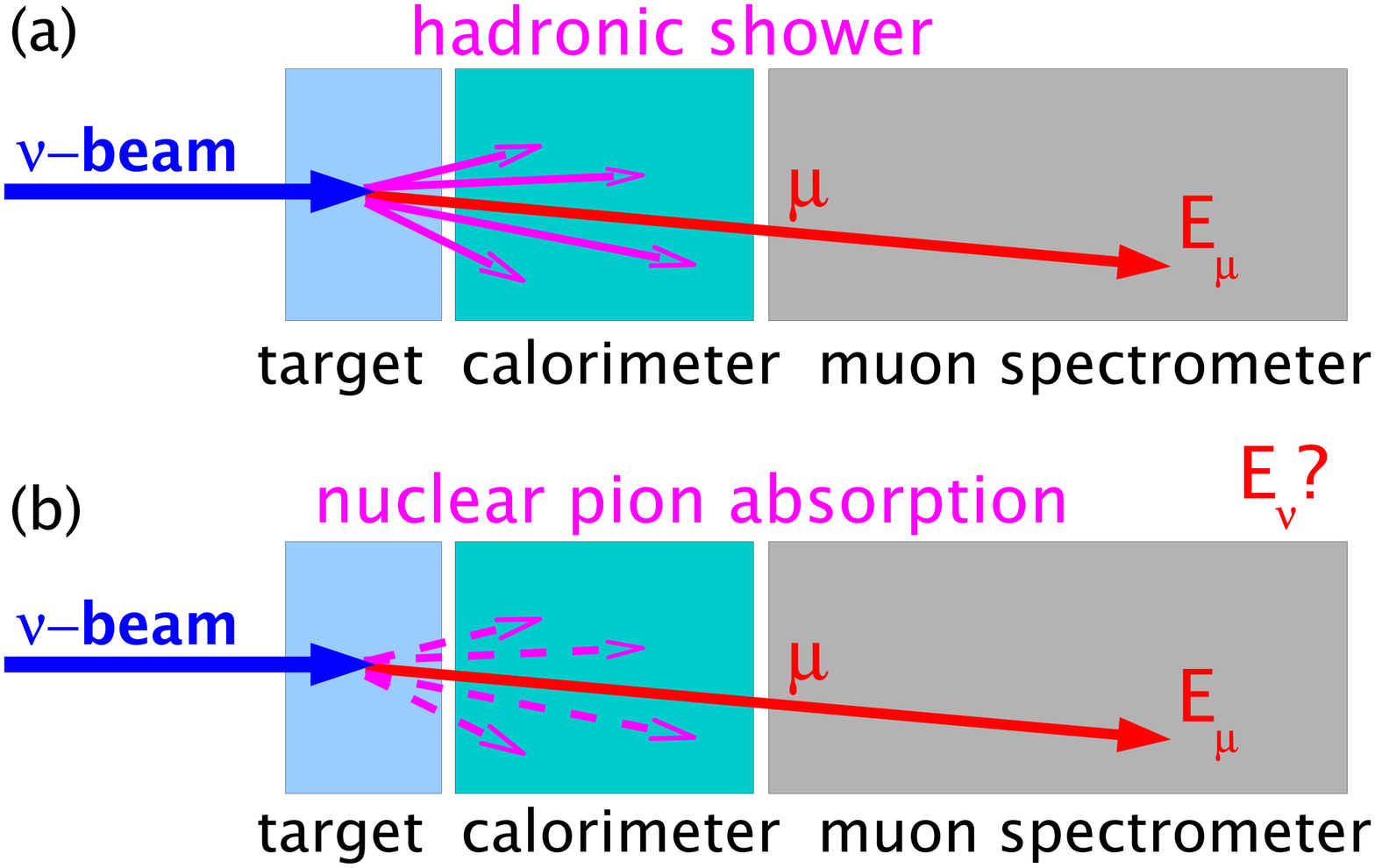}
\includegraphics[height=1.5in]{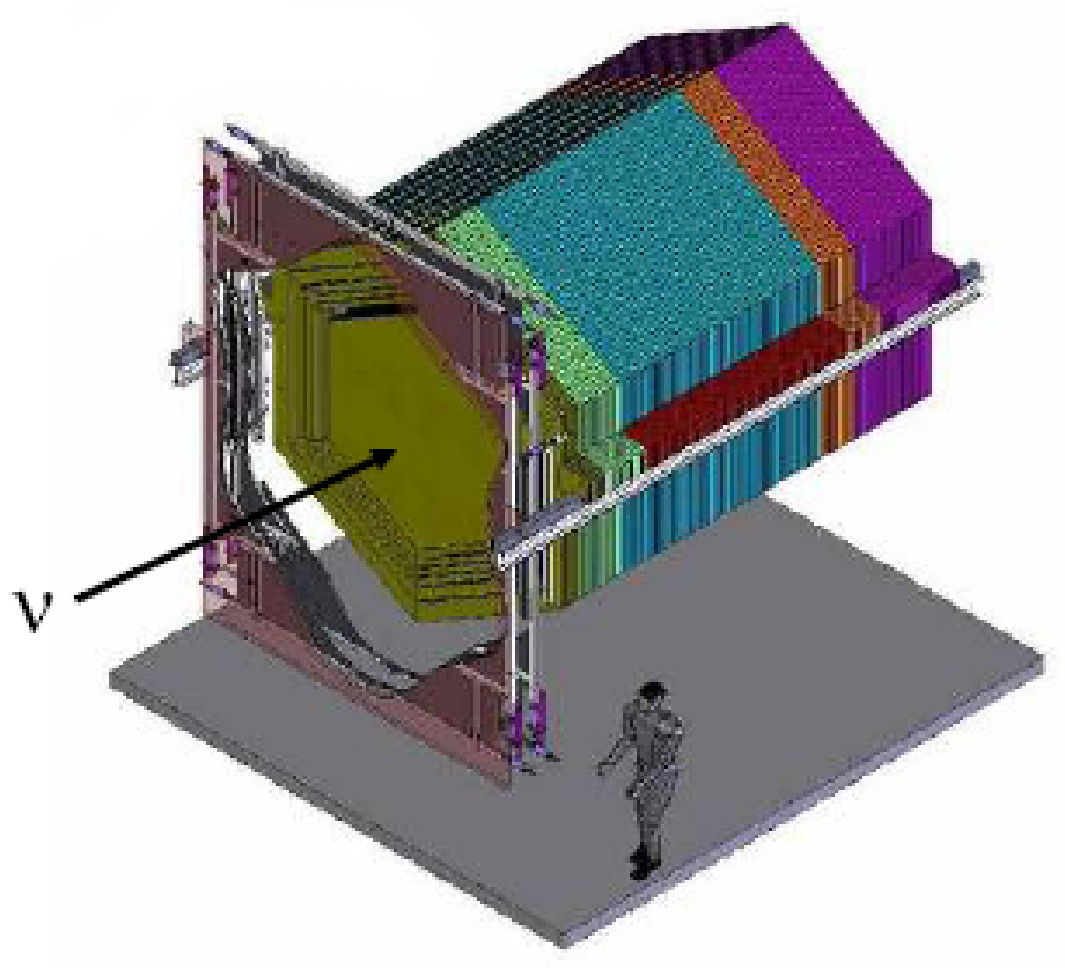}
\includegraphics[height=1.5in]{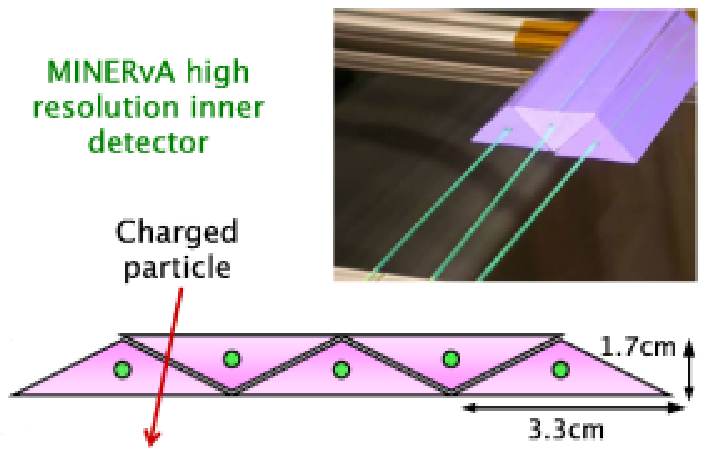}
\caption{
(Left) calorimetric energy reconstruction. 
The detector can be separated into 3 part, target, calorimeter, and muon spectrometer. 
(a) is the ideal situation, but often hadronic showers are missed, 
for example by nuclear pion absorption like (b), and give incorrect neutrino energy. 
(Middle) MINERvA detector. Front planes are target region, 
where MINERvA has ability to switch nuclear targets. 
The interior of the detector consists of 
a high resolution organic plastic scintillation tracker, 
and the outside is a magnetized calorimeter. 
(Right) picture of MINERvA inner detector scintillation-bar and schematic view. 
It consists of plastic organic scintillator with wavelength shifting fibers.  
The array of triangular bars have high resolution 
by the extraction of particle tracks from the amount of shared light 
by each scintillation-bar.}
\label{fig:calor}
\end{figure}

Neutrino energy misreconstruction happens, for example, when 
hadronic showers are absorbed by nuclei (Fig.~\ref{fig:calor}, left). 
This is important for precise $\nu_\mu$ disappearance measurements 
by MINOS, where steel is used as a target but no reliable 
pion absorption measurements are available. 
The future Main Injector Experiment for $\nu$-A (MINERvA) 
has the ability to switch its target and they plan to study nuclear effects 
(Fig.~\ref{fig:calor}, middle and left) 
as well as various physics topics from quasi-elastic to DIS~\cite{MINERvA}. 
Their measurements will significantly reduce the uncertainties on $\De m_{23}^2$ 
coming from nuclear cross section modeling in MINOS~\cite{MINERvA}.

\section{Background channel \label{sec:bkgd}}


Since T2K uses water \v{C}erenkov detector ``Super-K'' as a far detector, 
the signal of $\th_{13}$, namely $\nu_e$ appearance 
is a single electron (Eq.~\ref{eq:nueCCQE}) 
because outgoing protons are below \v{C}erenkov threshold in most cases and therefore invisible.

The notorious background for this signal is the neutral 
current $\pi^{\circ}$ (NC$\pi^{\circ}$) interaction,
\begin{eqnarray}
\nu_\mu+N \to \nu_\mu+N+\pi^{o}.
\label{eq:NCpi0}
\end{eqnarray}

Although $\pi^{\circ}$ decays to two photons, there are various reasons to miss one of them, 
for example, two photons overlap, or one photon is boosted to low energy 
below threshold. The precise prediction of this channel is critical for any $\nu_e$ 
appearance experiments. K2K measured the NC$\pi^{\circ}$ rate using 1KT detector~\cite{K2K_ncpi0}. 

\begin{figure}
\vskip 0.0cm
\hskip 3.0cm
\includegraphics[height=2.0in]{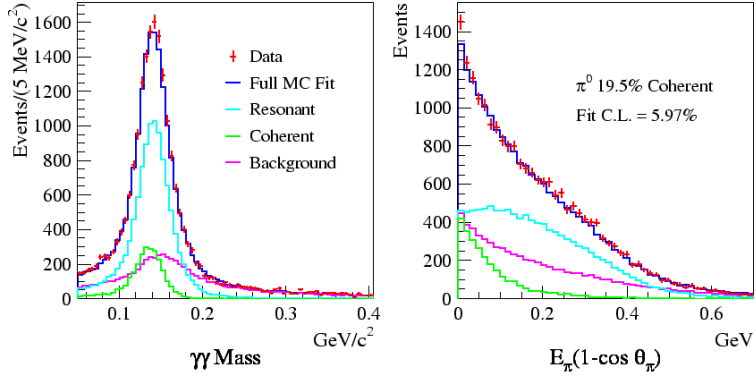}
\caption{
(Left) reconstructed $\pi^{\circ}$ mass peak for MiniBooNE after the correction of 
$\pi^{\circ} $ kinematics and coherent $\pi^{\circ}$ fraction. 
(Right) data-simulation comparison of one of kinematic variable. 
The template fit obtains a $19.5\%$ coherent fraction 
(coherent events are sharply peaked in the forward direction, 
{\it i.e.}, low $E_{\pi}(1-cos\th_\pi)$). 
}
\label{fig:MB_ncpi0}
\end{figure}

Recently, the MiniBooNE experiment made an {\it in-situ} measurement of 
NC$\pi^{\circ}$ production on mineral oil 
which was used to predict background processes more precisely for their 
$\nu_e$ appearance search~\cite{MB_ncpi0}.
Even though the underlying source of the $\pi^{\circ}$ may not be known, 
({\it i.e.}, actual resonance model to create the $\pi^{\circ}$ is not clear), 
the difference between the observed and predicted kinematic distribution of 
$\pi^{\circ}$'s can be used to correct the rate of $\pi^{\circ}$ events 
that are misclassified as $\nu_e$ signal events. 
Since the loss of a photon in the $\pi^{\circ}$ decay is mostly a kinematic effect, 
once correct $\pi^{\circ}$ production kinematics are obtained from the data, 
it is easy to calculate the distribution of $\pi^{\circ}$ where one photon is missed.
Left plot of Fig.~\ref{fig:MB_ncpi0} shows data-simulation comparisons for pion mass peak. 
After the correction, their simulation precisely predicts all observed aspects 
of NC$\pi^{\circ}$ events. 
The right plot of Fig.~\ref{fig:MB_ncpi0} shows a kinematic distribution. 

This result triggered another interest. This plot clearly shows the existence of 
NC coherent pion production.  
However, the K2K experiment saw no evidence for CC coherent pion production 
at similar energies~\cite{K2K_ccpip}. 
Since a coherently produced pion has very different kinematics, 
understanding of this rate is important. 
Again, further analysis of K2K, MiniBooNE, SciBooNE, MINOS, 
and MINERvA will shed light on this in the near future.

\vspace{0.5cm}
The fine-grained MINERvA detector will provide critical input for NOvA. 
Although high statistics data from K2K, MiniBooNE, 
and SciBooNE will be available, 
backgrounds of $\nu_e$ appearance search around $\sim 2$ GeV is only 
effectively accessible by MINERvA experiments. 
We are expecting negligible cross section error on $sin^22\th_{13}$ 
from NOvA after precise CC and NC measurements from MINERvA~\cite{MINERvA}.

\section{Conclusions}

The goal of next-generation long baseline accelerator-based neutrino oscillation experiments 
is to measure a $\nu_e$ appearance signal. 
The cross section errors arise from (1) misreconstruction of neutrino energy and 
(2) incorrect background predictions.
The inputs from current and future neutrino cross section measurements 
are critical to the success of future oscillation experiments, such as T2K and NOvA.

\end{document}